\documentclass[aps,prl,preprint,showpacs]{revtex4}
\usepackage{graphicx}

\usepackage{amsmath,amssymb}
\begin{document}
\title{Effective temperature of active matter}
\author{Davide Loi}
\affiliation{
European Synchrotron Radiation Facility, BP 220, F-38043 Grenoble, 
France}
\author{Stefano Mossa}
\affiliation{
European Synchrotron Radiation Facility, BP 220, F-38043 Grenoble, 
France}
\author{Leticia F. Cugliandolo}
\affiliation{Universit\'e Pierre et Marie Curie -- Paris VI, LPTHE UMR 7589,
4 Place Jussieu,  75252 Paris Cedex 05, France}
\date{\today}
\begin{abstract}
We follow the dynamics of an ensemble of interacting self-propelled
motorized particles in contact with an equilibrated thermal 
bath. We find that the fluctuation-dissipation relation allows for 
the definition of an effective temperature that is compatible with 
the results obtained using a tracer particle as a thermometer.
The effective temperature takes a value which is higher than 
the temperature of the bath and it is continuously controlled by 
the motor intensity.
\end{abstract}
\maketitle

Active matter is driven out of equilibrium by internal or external
energy sources. Its constituents absorb energy from their environment
or from internal fuel tanks and dissipate it by carrying out internal
movements that lead to translational or rotational motion. 
A typical example can be found in eukariotes, which are cells
organized into complex assemblies by internal structures.
In these cells many processes involve the cytoskeleton~\cite{cytoskeleton}, 
the cellular scaffolding, that is formed by a network of long polar 
filaments interacting through molecular motors.
The latter exert stresses which deform the former and regulate the network 
dynamics. These structures exhibit a rich variety of viscoelastic properties; 
by rearranging their structure they change from plastic/fluid to 
elastic/glassy phases and {\it vice versa}. Another celebrated example of 
active matter are self-propelled particle assemblies in bacterial 
colonies~\cite{bacteria}.

Here, we address the following fundamental question: which thermodynamic 
concepts, if any, can be applied to active matter? We focus on the 
definition of an {\it effective temperature}, $T_{eff}$, through the 
comparison of induced and spontaneous dynamic fluctuations~\cite{Teff}. 
The notion of such a $T_{eff}$ was introduced in the study of (passive) 
glassy systems, that is to say, macroscopic objects with a sluggish 
relaxation that takes them to an asymptotic small-entropy production regime 
which is, though, still far from equilibrium. In these systems $T_{eff}$ 
is a good thermodynamic concept. It admits a microcanonical-like definition 
in which only asymptotically dynamically accessible states contribute to the 
entropy~\cite{Makse}. $T_{eff}$ can be measured by a fine-tuned thermometer, 
it satisfies a zero-th law in the sense that it takes the same value for all 
observables evolving in the same time-scale, and heat flows in the direction 
of negative effective temperature differences~\cite{Teff}.  
Moreover, it appears in the modification of fluctuation theorems when applied 
to systems that cannot equilibrate when let freely evolve~\cite{Zamponi}.  
These features were confirmed with numerical simulations of a variety of 
realistic glassy models~\cite{Binder-Kob} and a number of groups are testing 
these ideas experimentally on colloidal suspensions~\cite{Teff-exp-coll}.
Theory suggests that the effective temperature should also be well-behaved 
in weakly driven systems -- in a small entropy production regime~\cite{Teff}. 
This fact was put to the test in gently sheared super-cooled liquids and 
glasses~\cite{Ludovic}, vibrated granular matter~\cite{Makse}, and moving 
vortex phases in superconductors~\cite{Kolton}. In biologically inspired 
problems the relevance of $T_{eff}$ was already stressed in studies of gene
networks~\cite{Wolynes1} and to reveal the active process in hair
bundles~\cite{Julicher} and in a simple three-component model system 
consisting of myosin II, actin filaments, and cross-linkers~\cite{mizuno}.
The modification of the rheological behavior in active filament solutions
was also interpreted in terms of a non-equilibrium temperature 
in~\cite{ziebert07}.

The role played by $T_{eff}$ in the stability of dynamic phases of
motorized particle systems was stressed by Shen and Wolynes~\cite{Wolynes2} 
with a variational analysis of the master equation of a similar model 
to the one we study here. In order to investigate the existence and 
properties of $T_{eff}$ in active matter we study a schematic statistical 
physical model: a system of $i=1,\dots,N$ massive, interacting and 
motorized spherical particles confined to a fixed volume, $V$, in $3d$. 
Spherical particles are indeed too simple for real applications in the 
context of cytoskeleton studies. In particular, they are allowed to 
freely move in space, with no bounds to a preexisting cytoskeletal 
filaments network. This choice is dictated by the need to highlight the 
effect of the non-equilibrium drive alone, with no additional effects 
related to a more complex internal structure. Similar behavior is 
anyhow expected for other geometries of the constituents. 
Particles are in contact with a thermal environment that is described by 
a random noise and a viscous drag. Neglecting hydrodynamics effects, 
the particles' velocities ${\bf v}_i$ thus evolve with the Langevin 
equation:
\begin{equation}
\label{eq:langevin}
 m_i\dot{\bf v}_i=
-\xi m_i {\bf v}_i
+ {\bf f}_i^s
+ {\bf f}^M_i+{\mathbf \eta}_i
\; . 
\end{equation}
Here, ${\mathbf \eta}_i$ is a Gaussian white noise representing thermal 
agitation with zero mean, and correlations
$\langle {\bf \eta}_i(t) {\bf \eta}_i(t') \rangle = 
2 \xi m_i T \delta(t-t')$ (we set $k_B=1$ henceforth). 
The term $-\xi m_i {\bf v}_i$ is the frictional force and $\xi$ 
the friction coefficient.
The overdamped character of the dynamics is taken into account by 
considering a large value $\xi=10$. The choice of Eq.~(\ref{eq:langevin}) 
in place of a position Langevin equation, where the inertia term  
$m_i\dot{\bf v}_i$ is dropped, allows for a more stable Verlet-like 
integration method and a more realistic description of short time 
dynamics~\cite{allentildesley}.

The systematic mechanical conservative force on particle $i$ due to all 
others is ${\bf f}^s_i \equiv \sum_{j=1}^N {\bf f}^s_{ij} = -\sum_{j=1}^N 
{\bf \nabla}_i U(r_{ij})$, with the $2n$-$n$ Lennard-Jones potential~\cite{Lekk}
$U(r_{ij})=4\epsilon [\left(\sigma/r_{ij}\right)^{2n}-\left(
\sigma/r_{ij}\right)^{n}]$. $r_{ij}$ is the interparticle distance.  
The short-range repulsion,
parametrized by $\sigma$, prevents the overlapping of the particles
and it is then a (soft) measure of the particle diameter.  The energy
parameter, $\epsilon$, describes the depth of the attraction.  The
mid-range attraction, $r_{Min}=\sigma 2^{-1/n}$, mimics the effective
cross-linking by linker proteins between elements in the
cytoskeleton. By choosing $n=18$ we get an adhesive soft-sphere
potential with $r_{Min} \ll \sigma$ as often used to describe
biomolecular assemblies~\cite{Wolynes2,Rosenbaum}. 
This case also presents the additional advantage that the region of 
the phase diagram where the metastable liquid phase is accessible 
in equilibrium is particularly extended. Again, we do not expect 
significant qualitative differences for other choices of $U$. 

Finally, motors apply a force ${\bf f}^M_i$ on each particle 
following a stochastic process that mimics the realistic chemical 
process. For simplicity, we use here a time series of isotropic kicks. 
During $\tau$ steps of the molecular dynamics trajectory independent 
forces are applied to a fraction of randomly chosen particles. 
The subset of propelled particles and the directions of the applied 
forces change at each power stroke of duration $\tau$. 
The strength of the force exerted on each particle is chosen to be 
a fraction of the mean mechanical force acting on the equivalent 
passive system, $\overline F=\frac{1}{N} \sum_{j=1}^N |{\bf f}_{ij}^s|$.  
The parameters  $\tau$ and $\overline F$ and the fraction of 
motorized particles are chosen consistently in order to induce an 
out-of-equilibrium steady state always remaining in linear-response 
conditions. We consider here {\it adamant} motors~\cite{Wolynes2} 
which act in such a way that the direction of the force is chosen at
random isotropically: their action is independent of the structural
rearrangements they induce. The case of {\it susceptible} motors,
which slow down when they drive the system uphill in the (free)-energy 
landscape, will be investigated in future work.

We integrated Eqs.~(\ref{eq:langevin}) numerically using Ermak's
algorithm~\cite{ermak78,allentildesley}. We henceforth use standard
Lennard-Jones units: length, energy and time-units are $\sigma$,
$\epsilon$ and $(m\sigma^2/\epsilon)^{\frac{1}{2}}$,
respectively~\cite{allentildesley}. In all our simulations we used
100 independent configurations with 500 particles. We have checked 
that larger systems have equivalent structural and dynamical behavior. 

\begin{figure}[t]
\vspace{0.3cm}
\includegraphics[width=0.70\textwidth]{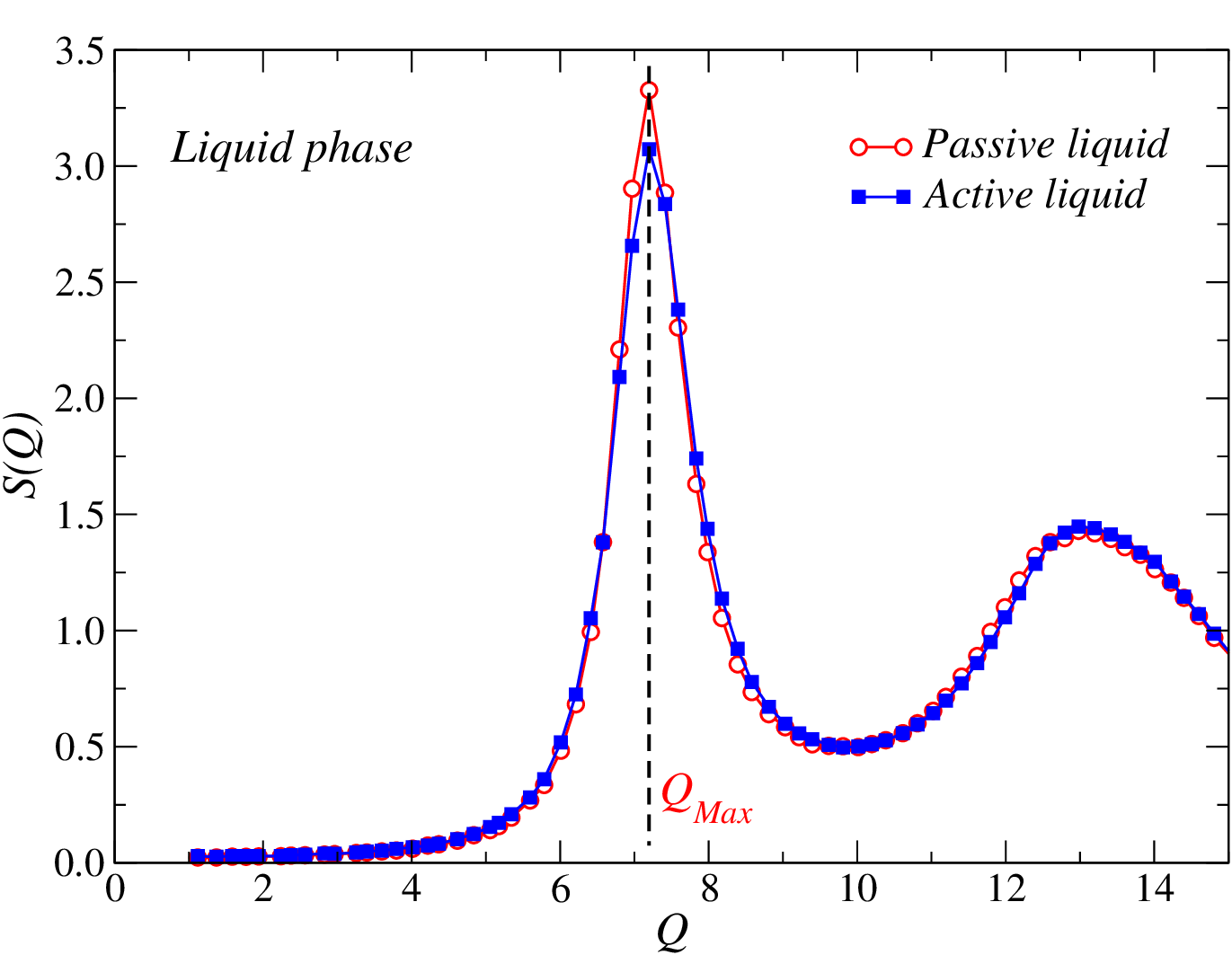}
\caption{\label{fig:structure} (Colour online)
The static structure factor, $S(Q)$, for a system in equilibrium and 
driven out of equilibrium by motors with $f^M=50\% \overline F$.}
\end{figure}

The passive system (motors switched off) is well characterized and
presents gas, liquid and crystalline phases depending on the bath
temperature and particle density~\cite{Lekk}. By tuning the intensity
and direction of the non-conservative forces we drive the system near
equilibrium (weak perturbation) or far from equilibrium (strong
perturbation).  We delay the description of the dynamic phase diagram
to a more detailed publication. Here we focus on the driven dynamics
of a system at conditions such that it is a liquid in equilibrium:
$T=0.8$ and $\rho=1$. 

We show data using a time-scale for the power strikes $\tau=5\times 10^3$, 
we apply the force to $10\%$ of the particles and the strength of the
non-conservative force, $f^M$, is indicated in each figure.
Other choices of these parameters give equivalent results.
We start by comparing the structure of active and passive matter.  
In both cases the structure factor, $S(\vec Q) = \langle \, \rho(\vec Q,t)
\rho^*(\vec Q,t) \, \rangle $, with $\rho(\vec Q,t)=N^{-1}
\sum_{i=1}^N e^{i\vec Q \vec r_i(t)}$ the Fourier transform of the
instantaneous density, becomes stationary and isotropic after a short
transient: $S(\vec Q,t)\to S(Q)$. In Fig.~\ref{fig:structure} we show
that $S(Q)$ for the passive liquid and the active system with
$f^M=50\% \overline F$ are practically identical, with a maximum
located at $Q_{Max}\sim 7.1$ in both cases.

\begin{figure}[t]
\vspace{0.3cm}
\includegraphics[width=0.70\textwidth]{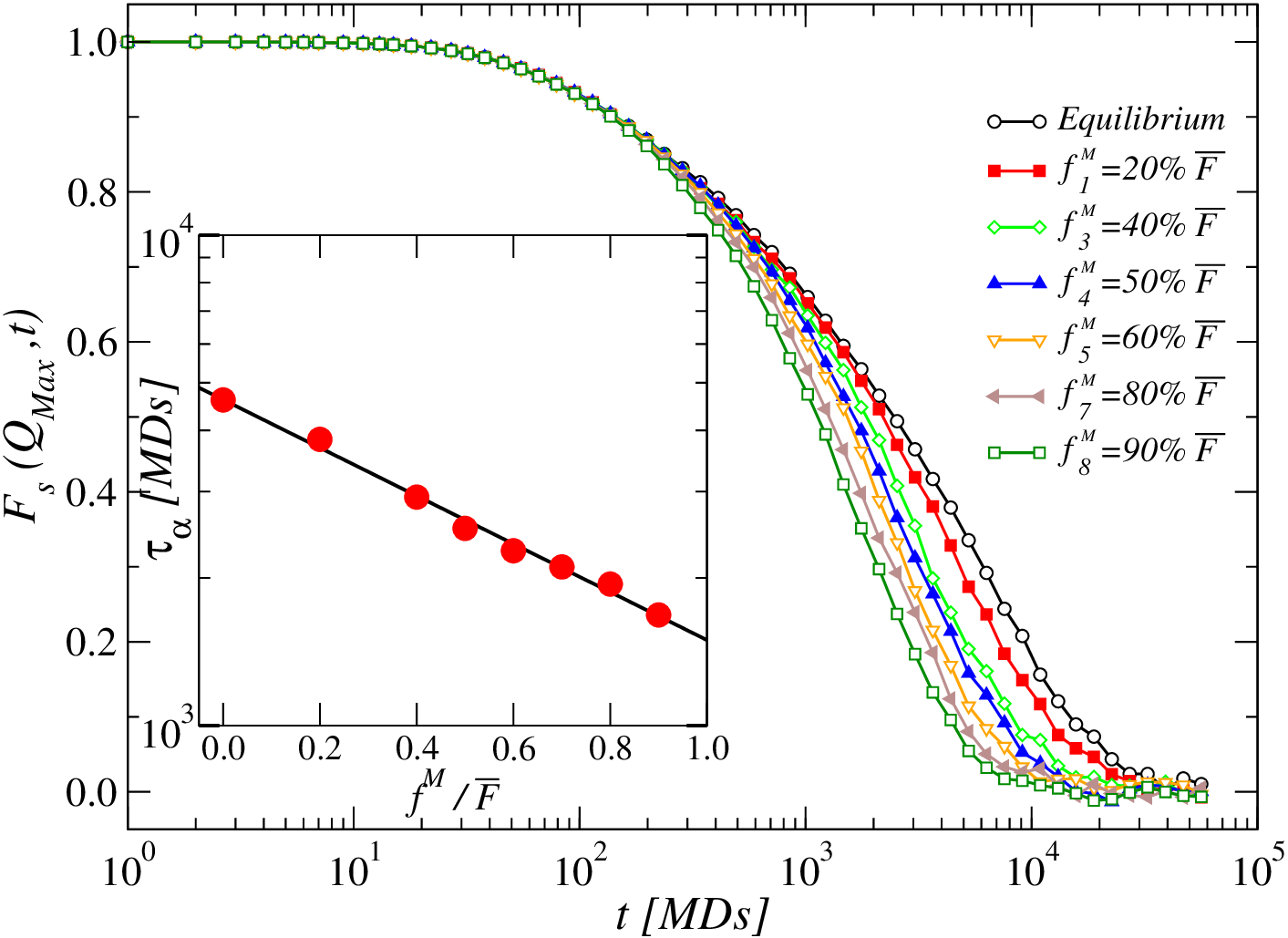}
\caption{\label{fig:incoherent} (Colour online)
Main panel: the incoherent intermediate scattering function at 
 $Q=Q_{Max}$, for motor force intensities given in the key. 
Data are shown as a function of simulation time measured in
molecular dynamics integration steps (MDs). 
Inset: the dependence of the $\alpha$ relaxation time on the strength of the 
motor forces with an exponential fit $\tau_\alpha\sim A \exp -(f^M/f^o)$ 
with $f^0\simeq 0.90$.}
\end{figure}

The relaxational dynamics of super-cooled liquids and glasses is
usually examined by monitoring the decay of the -- possibly two-time
dependent -- incoherent (one-particle) intermediate scattering
function $F_s(\vec Q,t,t_w)=N^{-1} \sum_{i=1}^N \langle \, e^{-i\vec
  Q[\vec r_i(t+t_w) -\vec r_i(t_w)]}\, \rangle $~\cite{Binder-Kob}
with $t_w$ the waiting-time measured after preparation and $t$ the
delay time between the total and the waiting times. In the liquid and
under the effect of the motors the systems reach an isotropic
stationary regime: $F_s$ only depends on $Q$ and $t$.
Figure~\ref{fig:incoherent} displays $F_s$ as a function of time-delay
in a log-linear scale for several force strengths ranging from zero
(equilibrium limit) to $90\% \overline F$. The decay is faster for
increasing force strengths though remains relatively slow for all
drives as demonstrated by the fact that $F_s$ decays to zero in a
logarithmic time-scale.  In the inset we show how the
$\alpha$-relaxation time, $\tau_\alpha\equiv
F_s(Q,\tau_\alpha)=e^{-1}$, decreases for increasing force strength in
a linear-log scale, similarly to the shear-thinning phenomenon.  The
line is $\tau_\alpha\sim A \exp -(f^M/f^o)$.  At each
applied force we find $\tau_\alpha \propto Q^{-2}$ (not shown).

\begin{figure}[t]
\vspace{0.3cm}
\includegraphics[width=0.70\textwidth]{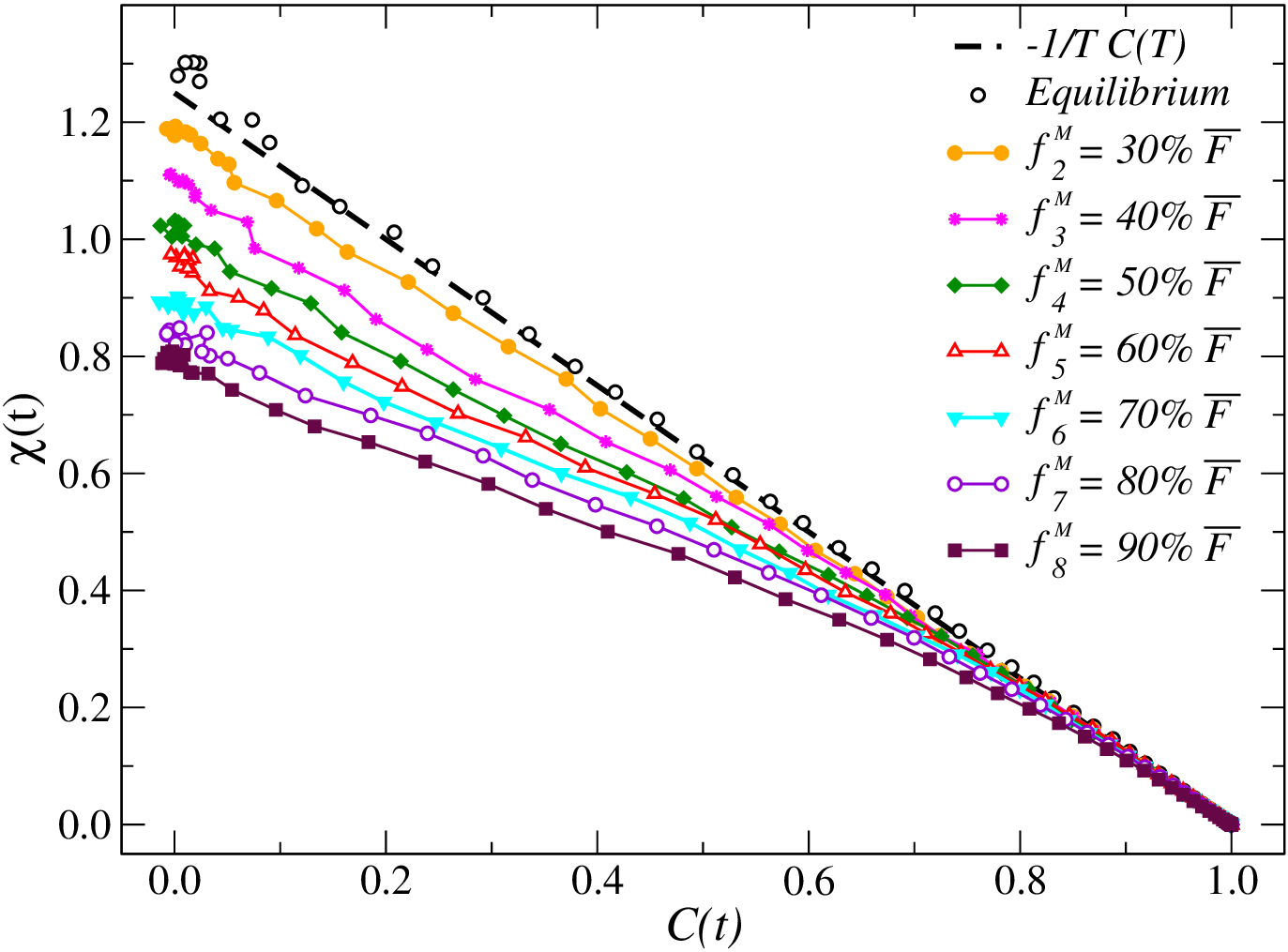}
\caption{\label{fig:fdt} (Colour online) Parametric representation of the 
fluctuation dissipation 
relation for passive and active matter. The strength of the applied 
forces are given in the key.}
\end{figure}

The equilibrium fluctuation-dissipation theorem (FDT) states that
the spontaneous fluctuations
are related to their associated induced
fluctuations by a model-independent formula
$T \chi(t,t_w) = [C(t=0,t_w)-C(t,t_w)]$.
$C$ is the connected correlation of two observables
$O_1$ and $O_2$, measured at times $t+t_w$ and $t_w$ 
respectively. $\chi$ is the linear response of the 
average of the observable $O_1$ measured at time 
$t+t_w$ to an infinitesimal perturbation that modifies the 
Hamiltonian as $H\to H-h O_2$ from $t_w$ to $t$:
$\chi(t,t_w) = \langle O_1^h(t+t_w)-O_1(t+t_w)\rangle /h$.
The choice of the appropriate observables $O_1$ and $O_2$
follows standard procedures and details about the calculations
can be found in Ref.~\cite{Ludovic}. Here, we recall that
in interacting particle problems~\cite{Ludovic} it is customary 
to use
$O_1(t_w) =\frac1{N} \sum_{i=1}^N
\epsilon_i e^{i\vec Q\vec r_i(t_w)}$, 
$O_2(t_w) = 
2 \sum_{i=1}^N \epsilon_i \cos[\vec Q\vec r_i(t_w)]
$, where the field $\epsilon_i=\pm 1$ with probability a half. 
With this choice, $C(t)=F_s(Q, t)$.
For each system configuration we averaged over 100 field realizations 
and considered $Q=Q_{Max}$. In Fig.~\ref{fig:fdt} we show the 
fluctuation-dissipation relation for passive and active matter.
The plots are parametric constructions of the integrated
linear response $\chi$ as a function of the correlation function using
$t$ as the parameter~\cite{Teff}. In the equilibrium case,
FDT holds and minus the inverse slope of
$\chi(C)$ is the temperature of the thermal bath, $T=0.8$ in this
case. In general, all curves join at $(C=1,\chi=0)$, as imposed by
normalization at $t=0$. The `initial' slope is determined by the
equilibrium FDT~\cite{Cudeku}.  At longer time-differences the curves
progressively depart from the equilibrium result to reach other
straight lines characterized by slopes that depend on the strength of
the motor forces. A time (or correlation) dependent effective
temperature is then defined as~\cite{Teff}
\begin{equation}
1/T_{eff}(C) = -d\chi(C)/dC
\; . 
\label{eq:Teff}
\end{equation}
Interestingly enough, soon after departure from the equilibrium form,
the curves approach a new straight line 
from which one extracts a single value of the effective
temperature -- within numerical accuracy~\cite{Houches}.

\begin{figure}[t]
\vspace{0.3cm}
\includegraphics[width=0.70\textwidth]{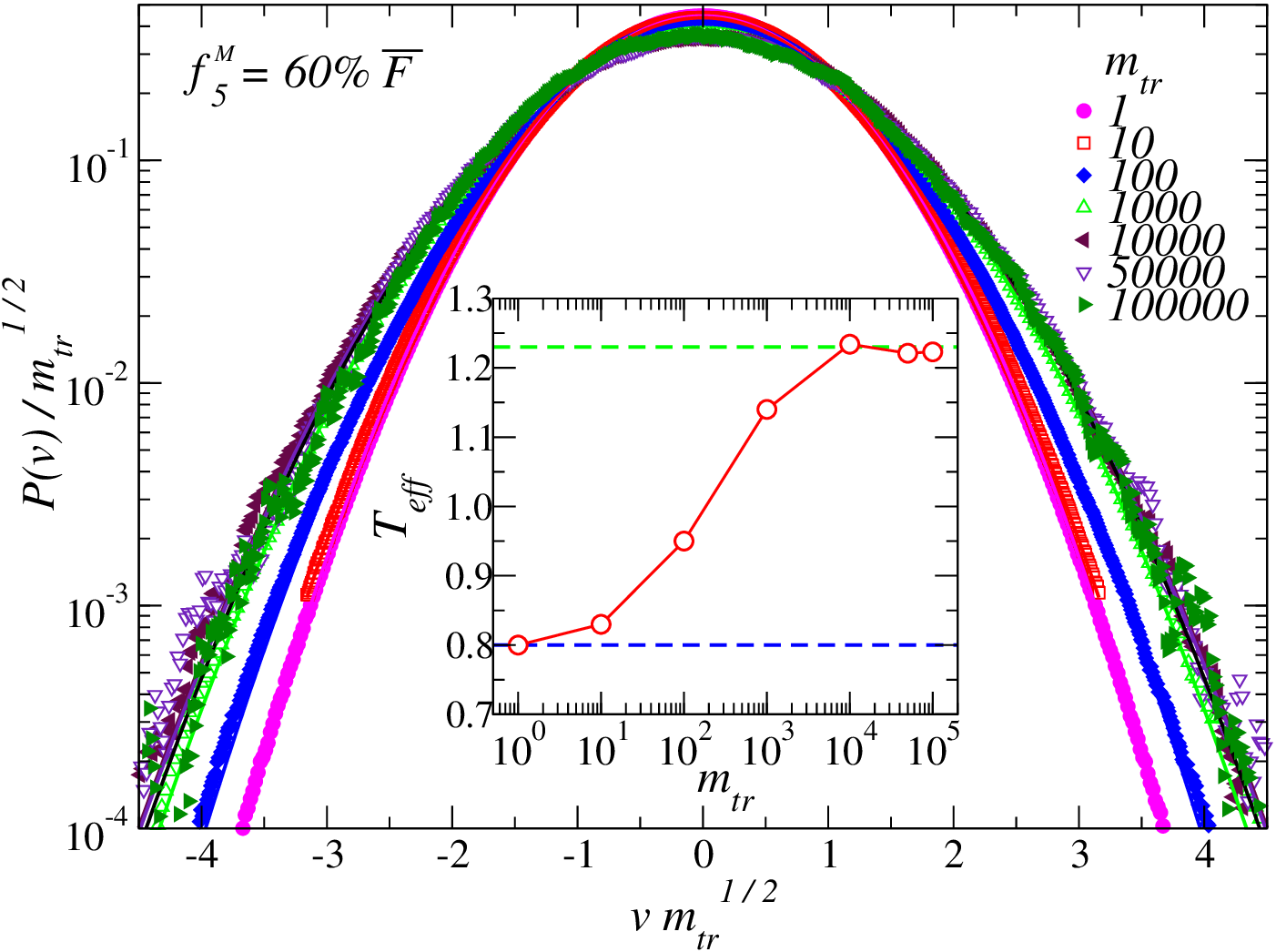}
\caption{\label{fig:tracer} (Colour online) Main panel: probability
distribution function of the velocity $v$ of different tracers with
masses $m_{tr}$ given in the key. The motor force intensity is $f^M=60\%
\overline F$. The full lines are fits to the Gaussian form
(\ref{eq:gaussian}). Inset: $T_{eff}$ as function of $m_{tr}$
as obtained from the Gaussian fits to the data-points in the main
panel.}
\end{figure}

The meaning of $T_{eff}$ as a temperature depends on it verifying a
number of conditions expected from such thermodynamic concept. In
particular, a temperature should be measurable with a thermometer. A
tracer particle with a long internal time-scale (proportional to the
square root of its mass)~\cite{Teff,Ludovic,Makse} acts as a
thermometer that couples to the long time-delay 
structural rearrangements (and not to the fast vibrations, $t\sim 0$,
that yield the ambient temperature, see Fig.~\ref{fig:fdt} and
\cite{Cudeku}).  We then coupled a tracer particle with mass $m_{tr}$
to the active matter {\it via} the same Lennard-Jones potential, so
that we do not modify the structure factor of the fluid.
(The tracer does not couple to the thermal bath.)  
In Fig.~\ref{fig:tracer} we show the
tracers' velocity distributions (for 10 independent tracers).  The
data points are superposed to 
\begin{equation}
P(v) = \sqrt{\frac{m_{tr}}{2\pi T_{eff}}} \; 
\exp\left(-\frac{m_{tr} v^2}{2T_{eff}}\right)
\; , 
\label{eq:gaussian}
\end{equation}
with $T_{eff}$ the only fitting parameter.  The reason why a
Maxwellian distribution applies is that the tracer behaves as a normal
system immersed in an environment made of active matter.  The inset
in Fig.~\ref{fig:tracer} displays the dependence of $T_{eff}$ on
$m_{tr}$ for one value of the energy pumping force. A very light
tracer basically follows the very fast -- high frequency -- dynamics
of its environment and thus measures the bath temperature (lower
horizontal dashed line). A heavier tracer feels the slower -- low
frequency -- structural relaxation and measures a higher
temperature. A sufficiently heavy tracer only follows the slow
dynamics of the sample and therefore measures the actual effective
temperature (upper horizontal dashed line), $T_{eff}(f^M=60\%
\overline F) \sim 1.23$.

\begin{figure}[t]
\vspace{0.3cm}
\includegraphics[width=0.70\textwidth]{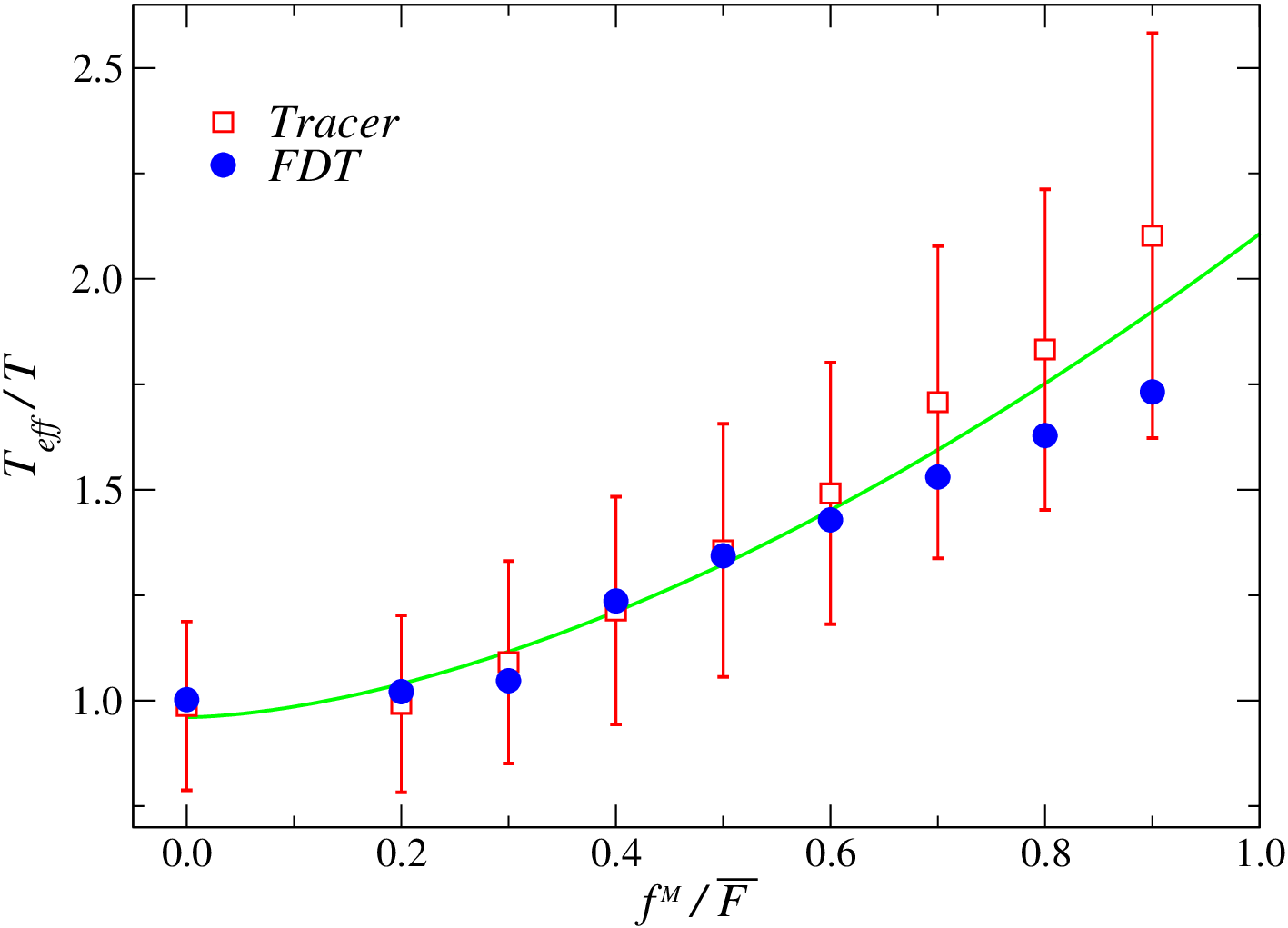}
\caption{\label{fig:comparison} (Colour online)
Effective temperature obtained from response-correlation and 
massive tracer calculations for different motor force intensities. 
}
\end{figure}

Finally, in Fig.~\ref{fig:comparison} we compare the values of
$T_{eff}$ derived from Eqs.~(\ref{eq:Teff}) and (\ref{eq:gaussian}).
The two measurements yield consistent results, with $T_{eff}$
increasing from $T$ at zero pump to about $2T$ when the strength of
the driving forces equals the one of the mechanical forces.
We also 
found that $T_{eff}$ decreases with decreasing $\tau$ (not shown). 

Summarizing, we showed that the notion of an effective temperature,
defined by studying the deviations from the equilibrium FDT out of
equilibrium can be applied to active matter. We demonstrated that 
the value of the effective temperature measured with a thermometer
realized by a tracer particle coincides -- within numerical accuracy
-- with the one obtained from the fluctuation-dissipation relation 
for all values of the pumping forces. For adamant motors $T_{eff}$ 
is larger or equal than the bath temperature and increases as a 
function of the pumping force. This finding is consistent with the 
intuitive idea that ascribes $T_{eff}$ to the degree of additional 
agitation caused by the non-conservative forces.
It will be interesting to study in detail the case of 
{\it susceptible} motors~\cite{Wolynes2} which slow down when driving 
the system uphill in the (free)-energy landscape. It is natural to 
expect that the effective temperature of such active matter would be 
{\it lower} than the one of the thermal bath. 
Finally, it is a challenge to derive the above results in the framework
of hydrodynamic theories of the kind proposed in~\cite{hydro}.

LFC is a member of Institut Universitaire de France.

\end{document}